\documentstyle[12pt]{article}
\newcommand{\mrm}[1]{\mbox{\rm #1}}
\newcommand{\half}{{1\over 2}}
\newcommand{\bla}{\hspace{1cm}}
\newcommand{\beq}{\begin{equation}}
\newcommand{\eeq}[1]{\label{#1}\end{equation}}
\newcommand{\eq}[1]{eq.~(\ref{#1})}

\newcommand{\db}{\not\hspace{-0.7ex}D\hspace{0.7ex}}
\def\titlepage{\clearpage%
\setcounter{footnote}{0}\setcounter{page}{1}%
\thispagestyle{empty}\pagestyle{plain}\pagenumbering{arabic}%
\kern1mm\begin{center}
\mbox{\large EUROPEAN ORGANIZATION FOR NUCLEAR RESEARCH}
\end{center}
\vskip5mm\normalsize}
\def\docnum#1{\hbox to \hsize{\hskip123mm\hbox{#1}\hss}}
\def\date#1{\hbox to \hsize{\hskip123mm\hbox{#1}\hss}}

\def\title#1{\vskip1em\begin{center}\Large\bf#1\end{center}\vskip2.5em}
\def\author#1{\vskip0.5em{ #1}\vskip0.5em}
\def\inst#1{\vskip0.5em{ #1}\vskip1.0em}
\def\abstract{\begin{center}{\bf Abstract}\end{center}\quotation}

\def\anotfoot#1#2{\vfill\noindent\underline{\hspace{6cm}}
\par\noindent #1) #2}
\begin{document}
\begin{titlepage}
\docnum{CERN--TH.6801/93}
\vspace{2cm}
\title{Masses, Mixings, Yukawa Couplings and their Symmetries}

\begin{center}
\author{Arcadi Santamaria$^{*)}$}
\inst{CERN, Geneva, Switzerland}
\end{center}
\vspace{2cm}
\begin{abstract}
We present a method to find the number of real and imaginary
observable parameters coming from the Yukawa sector in an arbitrary gauge
theory. The method leads naturally to a
classification of Yukawa couplings according to their symmetries
and suggests a new parametrization of masses and mixings that is
useful to study the behaviour of Yukawa couplings under the
renormalization group. We apply it to some examples based on the Standard
Model with Yukawa couplings obeying various chiral symmetries. We also
show how our method of parameter counting can be used in some
models with an enlarged leptonic sector.
 \end{abstract}
\vspace{2cm}
\vfill
CERN-TH.6801/93\\
February 1993
\anotfoot{*}{On leave of absence from Departament de
F\'{\i}sica Te\`orica,
Universitat de Val\`encia and IFIC, Val\`encia, Spain.}
\end{titlepage}

The most difficult part to understand of theories with spontaneously
broken gauge theories is, together with the Higgs potential, the Yukawa
sector.
Its origin is unclear,
and probably related to the origin of the spontaneous
breaking of the symmetries itself. It contains a large number of free
parameters that must be adjusted by hand to obtain a
realistic spectrum of
particles and mixings. From the 17 parameters needed
to describe the electro-weak theory, 13 come from the Yukawa sector.
In addition, the Yukawa couplings contain a lot of
spurious parameters that cannot be observed because they can be removed
from the Lagrangian with an
appropriate choice of phases and mixings for the
fermion fields. Thus, a simple question such
as how many parameters are needed
to describe a theory with Yukawa
couplings cannot be answered at first sight.
We cannot even say whether the couplings
are real or not and so if the theory
conserves CP or not. This represents a great
complication when one tries to
understand mass matrices in the Standard Model. In the last
years this
subject has received a lot of attention
(for some popular mass-matrix
models, see for example
\cite{Fri79,Ste83,Bar87,Giu92}).
Usually people make ans\"atze about
a particular form of mass matrices and
try to extract from them a reasonable
spectrum of masses and mixings. However,
completely different forms of mass
matrices may be equivalent in their physical content.
This has led to some
confusion in the past. Moreover, some symmetries actually
present in the Yukawa
sector may be hidden when one does not work in the appropriate basis. For
those reasons, an analysis based on symmetry should be superior and could
clarify some of the issues present in particular forms of mass matrices.
It would also  allow a clear understanding of the number
of parameters needed
to describe the model.

The standard method to calculate the number of parameters
needed to describe
the Yukawa sector of
a spontaneously broken gauge theory is just by construction.
After symmetry
breaking one diagonalizes the mass matrices of all fields and
after that one
uses the freedom in the phase definition of the fields to reabsorb
as many
phases as possible. The remaining parameters in the Lagrangian are the
physical parameters.  As we know, the method works very well for the
Standard
Model. However, as the complication of the Yukawa sector increases, the
method becomes  intricate and can lead very easily to error;
 the question
of CP-violation, for example,  can then be difficult to answer.
 In addition, in many cases, phases
and mixings can be moved from  the gauge couplings to the Yukawa sector
or even included in a redefinition of the fields \cite{BP83,BP86}, which
leads to further confusion.

Here we
propose a method, based on symmetry, to perform the
counting before spontaneous symmetry breaking and
without performing the explicit diagonalization of the
mass matrices. The method clarifies the structure of the manifold of the
parameters that are
needed to describe the Yukawa sector of a gauge
 theory and suggests a
parametrization that is
useful when one wants to study the behaviour of the Yukawa
couplings under the renormalization group.  Basically the method consists
in the study of the symmetries of both the full and
the Yukawa Lagrangians. The number of parameters needed to describe
the Yukawa sector comes from the balance between the number of parameters
contained in the Yukawa matrices and the number of symmetries broken by
the Yukawa couplings.
As an example,
we will start with the Standard Model with the most general
set of Yukawa couplings. We will then impose to the Standard Model a
set of chiral symmetries
that could reduce the number of parameters and see
how the method can be used in those cases. Finally we will apply the
 method to more involved models based on the same
 gauge group but different particle content.

The number of parameters needed to describe the Yukawa sector
of a  theory
is obviously
the same before and after spontaneous symmetry breaking\footnote{For the
moment, we leave aside the case of spontaneous breakdown of CP, in which
some parameters can be moved from the Higgs
potential to the Yukawa sector.}.
Thus, we will start with the hadronic part of the electro-weak
Lagrangian before symmetry breaking
\beq
{\cal L } = i \overline{Q_L} \db Q_L
            +i \overline{u_R} \db u_R
            +i \overline{d_R} \db d_R
            + (\overline{Q_L} Y_u u_R \varphi
            + \overline{Q_L} Y_d d_R \tilde{\varphi}
            + \mrm{h.c.}) \   .
\eeq{lagr}
Here $Q_L$, $u_R$ and $d_R$ are the standard quark fields, left-handed
doublet, u-singlet and d-singlet respectively.
If we assume $n$ generations
they are $n$-column vectors in generation space;
$Y_u$ and $Y_d$ are the Yukawa couplings represented by
$n\times n$ complex arbitrary matrices;
$ \db $$\equiv \gamma^\mu D_\mu$ where $ D_\mu$ is the Standard Model
electro-weak covariant derivative.
Finally $\varphi$ is the Higgs doublet and
$\tilde{\varphi}\equiv i\tau_2 \varphi^*$.

If $Y_u=Y_d=0$, the Lagrangian is obviously invariant under the
following chiral symmetries:
\beq
Q_L \rightarrow V_Q Q_L\bla
u_R \rightarrow V_u u_R\bla
d_R \rightarrow V_d d_R\   ,
\eeq{trcamps}
where, $V_Q$, $V_u$, and $V_d$ are $n\times n$ unitary matrices acting on
flavour space.
The Yukawa couplings break explicitly these symmetries, but in
a very particular way.  In fact, if we let the Yukawa couplings transform
as follows
\beq
Y_u \rightarrow V_Q Y_u V_u^\dagger\bla
Y_d \rightarrow V_Q Y_d V_d^\dagger\  ,
\eeq{trmatrius}
the Lagrangian of \eq{lagr} is still invariant under
the combined action of
the transformations in eqs. (\ref{trcamps}) and (\ref{trmatrius}).
It is also easy to
see that not all the symmetries in \eq{trcamps} are broken by the Yukawa
couplings. Indeed if we choose $V_Q=V_u=V_d= e^{i\alpha}$, the Yukawa
Lagrangian remains invariant. This is nothing else than baryon number
conservation.

Equation (\ref{trmatrius}) defines an equivalence relation
\beq
(Y_u,Y_d) \Leftrightarrow (Y'_u,Y'_d)=
(V_Q Y_u V_u^\dagger,\ V_Q Y_d V_d^\dagger)\  .
\eeq{releq}
The Lagrangians with couplings $(Y_u,Y_d)$ and couplings $(Y'_u,Y'_d)$
are completely equivalent. Thus, counting how many parameters are needed
to describe masses, mixings, and Yukawa couplings,
 is the same as counting
how many equivalent classes there
are with respect to the equivalence relation of
\eq{releq}.

This problem has some similarity with the problem of spontaneous symmetry
breaking by the Higgs mechanism. There, one has a group $G$ and a
representation $\varphi$ of Higgses whose vacuum expectation value breaks
the group $G$ down into
the group $G' \subset G$ and one wants to know the
number of physical Higgses.
As we know, the number of physical Higgs degrees of
freedom $N_{\varphi_{phys}}$ is
\beq
N_{\varphi_{phys}} = N_\varphi-N_{Goldstone}\ .
\eeq{nhiggs}
Here $N_\varphi$ is the number of degrees of freedom in the
Higgs representation $\varphi$ and $N_{Goldstone}$ is the number of
Goldstone bosons that appear after spontaneous symmetry breaking. It
is equal to the number of broken generators of $G$,
which is the number of generators of the full group $G$
minus the number of generators of the unbroken subgroup $G'$
($N_{Goldstone}=N_G-N_{G'}$).
We have a similar
situation here. A chiral symmetry $G$ is broken explicitly by the
Yukawa sector $Y$ to a group $G'$.  Then, only the broken part of
$G$ can
 be used to absorb parameters from $Y$. Following \eq{nhiggs}
we could write
\beq
N_{Y_{phys}} = N_Y-N_G+N_{G'} \ .
\eeq{eqcount}
 In the Standard Model, as follows from \eq{trcamps},  the group is
$G=U(n)_Q \otimes U(n)_u \otimes U(n)_d$
and the Yukawa couplings $Y=(Y_u, Y_d)$ are two general $n\times n$
complex matrices transforming under the group $G$ as \eq{releq}.
Here the factors $U(n)$ denote the full unitary group $U(n)=SU(n)\otimes
U(1)$. As discussed previously,
for the most general Yukawa couplings, the only subgroup of $G$ that
leaves the Yukawa couplings invariant is a $U(1)_B$ such that
$V_Q=V_u=V_d= e^{i\alpha}$.

After \eq{eqcount}, counting of parameters is trivial. Taking into account
that an arbitrary complex
matrix of dimension $n$ contains $n^2$
moduli and $n^2$ phases and that a $U(n)$
matrix
contains $n(n-1)/2$ moduli and $n(n+1)/2$ phases, we find that $Y_{phys}$
can be expressed in terms of
\beq
\begin{array}{||c|c|c||}\hline
\mrm{couplings and symmetries} & \mrm{moduli} & \mrm{phases} \\ \hline
(Y_u , Y_d)  &  2n^2 & 2n^2 \\ \hline
U(n)_Q \otimes U(n)_u\otimes U(n)_d & -3n(n-1)/2 & -3n(n+1)/2 \\ \hline
U(1)_B & 0 & 1 \\ \hline
 Y_{phys} &2n+ n(n-1)/2  & (n-2)(n-1)/2 \\ \hline
\end{array}\  .
\eeq{compte}
 In particular if $n=3$ we have $6+3$ moduli and $1$ phase.
After symmetry breaking, some of the moduli are related
to masses and other to mixings or remaining Yukawa couplings.
To distinguish
among them we can
repeat the argumentation but using the completely broken
theory with all the heavy  and Higgs bosons decoupled. In the case of the
Standard Model,
what remains is a model with the standard fermion content but
only QED (and QCD if included in the analysis)\footnote{We have
to keep all the
exact  symmetries. Fields with different charges never mix.}.
The resulting Lagrangian is
\beq
{\cal L } = i \overline{u_L} \db u_L
            +i \overline{u_R} \db u_R
            +i \overline{d_L} \db d_L
            +i \overline{d_R} \db d_R
            + (\overline{u_L} M_u u_R
            + \overline{d_L} M_d d_R +\mrm{h.c}) \ .
\eeq{simlagr}
Here $\db$ is the standard QED covariant derivative and
$M_u$ and $M_d$ are
the quark mass matrices related to the Yukawa couplings by the vacuum
expectation value of the Higgs scalar.

Except for the mass terms, the Lagrangian of \eq{simlagr}
is invariant under the following symmetries
\beq
u_L \rightarrow V_{u_L} u_L\bla
u_R \rightarrow V_{u_R} u_R\bla
d_L \rightarrow V_{d_L} d_L\bla
d_R \rightarrow V_{d_R} d_R\  .
\eeq{simtrcamps}
Using similar arguments as above, but now taking into account that the
full Lagrangian has  separate flavour conservation for $u$-type and
$d$-type quarks,  i.e.
$G=U(n)_{uR}\otimes U(n)_{uL}\otimes U(n)_{dR}\otimes U(n)_{dL}$
and $G'= (U(1)_u)^n\otimes (U(1)_d)^n$, we find that the
Lagrangian (\ref{simlagr}) gives rise to only $2n$ masses. Of course
this is obvious since the symmetries in \eq{simtrcamps} are enough to
diagonalize the two mass matrices completely.
Thus, from the $2n+ n(n-1)/2$ moduli present in the
Yukawa sector of the Standard Model, $2n$ correspond to masses
and $n(n-1)/2$ to mixings, as expected \cite{KM73} (see also \cite{BC86}).

 Yukawa couplings related by \eq{releq} are in the same class of
 equivalence. To find  a parametrization of the physical Yukawa sector
 ve have to characterize the equivalent classes, and this can be done by
 taking one element of  each equivalence class.  We can use a
 $V_Q$ and a $V_u$ rotation to diagonalize the matrix $Y_u$, $V_Q Y_u
 V_u^\dagger=D_u$. After that, since any arbitrary complex matrix
 can be written as a Hermitian matrix times a unitary matrix,
 we can use a $V_d$
 transformation to write $Y_d$ as a positive-definite Hermitian matrix.
 Thus $(Y_u,Y_d) \rightarrow (D_u, H_d)$, with $D_u$ a positive-definite
 diagonal matrix and $H_u$ a positive-definite Hermitian matrix.  But,
 $D_u$ is still invariant under transformations like $D_u\rightarrow K
 D_u K^\dagger$, with $K$ a diagonal matrix of phases, while $H_d$ is not.
 This means that we can use $K$ to absorb
 phases from $H_u$. In fact we can chose $K$ such that the
 next-to-diagonal
 diagonal elements of $H_d$ be real and positive. The only
 remaining symmetry is just $U(1)_B$ of baryon number, as expected.
 Thus, without loss of generality we can represent the physical
 Yukawa couplings of the Standard Model by a positive-definite diagonal
 matrix $D_u$ for the $u$-type quarks and a
 positive-definite Hermitian matrix
 $\hat{H}_d$ with the next-to-diagonal diagonal elements of $\hat{H}_d$
 real and positive:
 \beq
 (Y_u,Y_d) \rightarrow  Y_{phys}=(D_u,\hat{H_d})\  .
  \eeq{paramet}
 One can easily check that the right-hand side
 of \eq{paramet} contains also
 $2n+n(n-1)/2$ moduli and $(n-1)(n-2)/2$ phases.
 Of course, one can further
 write $\hat{H_d}$ in terms of a {\it Kobayashi-Maskawa}-type matrix and
 a diagonal matrix of masses.

 Since the full Lagrangian is invariant under
 the combined action of transformations (\ref{trcamps}) and
 (\ref{trmatrius}) it is easy to show that the renormalization group
 equations for the Yukawa couplings must be
 covariant with respect to the set of
 transformations (\ref{trmatrius}). This guarantees that the
 renormalization group equations  can be written only in
 terms of the parameters in the right-hand side of \eq{paramet}. This
 way we can obtain a set of renormalization group equations with all
 the unphysical parameters removed\footnote{A similar approach has
 been followed by \cite{Jar85,Bab87,OP89,OP91}}.

The method we just explained  for counting  the physical parameters
 works very well in the
Standard Model with the most general Yukawa couplings. However,  in the
form we have presented it, it is not suitable when some of the masses are
zero or there are some symmetries among them. We want to generalize the
method  to those cases. We will
impose that the only allowed reductions of parameters are those that
are protected by some symmetry.
 For instance, suppose we want to
make the first $n-1$ $d$-type quarks (in $n$ generations) exactly
massless.
We could impose an additional chiral symmetry such as
\beq
Q_L\rightarrow Q_L \bla u_R \rightarrow u_R\bla
d_R \rightarrow
\left(
\begin{array}{cc}
U & 0 \\
0 & 1
\end{array} \right) d_R\  ,
\eeq{bova}
where $U$ is a $(n-1)\times(n-1)$ general unitary matrix acting only
on the first $n-1$ generations of $d_R$ quarks. Clearly the only way
to make the Yukawa Lagrangian invariant under this symmetry is for
\beq
Y_u = \mrm{arbitrary}\bla  Y_d = \left(\begin{array}{ccccc}
0 & \cdot & \cdot & 0 & \alpha_1 \\
0 & \cdot & \cdot & 0 & \alpha_2 \\
\cdot & \cdot & \cdot & \cdot & \cdot \\
\cdot & \cdot & \cdot & \cdot & \cdot \\
0 & \cdot & \cdot & 0 & \alpha_{n}
\end{array}\right)\ .
\eeq{rescoup1}
Then, the Yukawa Lagrangian is only invariant
under $U(n-1)\otimes U(1)_B$.
Na\"{\i}ve counting starting from the full group
$G=U(n)_Q\otimes U(n)_u\otimes U(n)_d$ does not work in this case,
because the couplings in \eq{rescoup1} are not a representation of
the full group $G$. Said in another way, general $G$ transformations
do no preserve the form of the Yukawa matrices in \eq{rescoup1} and,
thus,
we cannot use the full group to absorb parameters from the Yukawa
sector. It is not difficult to see that the only subgroup of $G$
that preserves the form of the Yukawa couplings in \eq{rescoup1}
is  $\hat{G}=U(n)_Q\otimes U(n)_u\otimes U(n-1)\otimes U(1)_B$.
Now we can use $\hat{G}$ to absorb parameters from the Yukawa sector
and the counting of physical parameters comes as follows
\beq
\begin{array}{||c|c|c||} \hline
\mrm{couplings and symmetries} & \mrm{moduli} & \mrm{phases} \\ \hline
(Y_u , Y_d)  &  n^2+n & n^2+n\\ \hline
U(n)_Q \otimes U(n)_u\otimes U(n-1)_d\otimes U(1)_B
 & -(n-1)(3n-2)/2 & -n(3n+1)/2-1 \\ \hline
U(n-1)\otimes U(1)_B & (n-1)(n-1)/2 & (n-1)n/2+1 \\ \hline
 Y_{phys} &n+1+(n-1) & 0 \\ \hline
\end{array} \  .
\eeq{compte2}
The model leads to $n$ massive $u$-type quarks, $1$ massive
$d$-type quark and
$n-1$ mixings. There is no CP-violating phase. This  result can be
explicitly checked by full diagonalization of the Yukawa sector.

Another extreme case can be obtained by imposing an axial $U(1)$ symmetry
on the $d$-quark:
\beq
Q_L\rightarrow Q_L \bla u_R \rightarrow u_R\bla
d_{iR} \rightarrow d_{iR}\bla i=2,\cdots,n\bla
d_{1R} \rightarrow e^{i\alpha} d_{1R} \ .
\eeq{trans1}
Then, the Yukawa couplings take the following form
\beq
Y_u = \mrm{arbitrary}\bla  Y_d = \left(\begin{array}{ccccc}
0 & \alpha_{1,1} &\cdot & \cdot & \alpha_{1,n-1} \\
0 & \alpha_{2,1} &\cdot & \cdot & \alpha_{2,n-1} \\
\cdot & \cdot & \cdot & \cdot & \cdot \\
\cdot & \cdot & \cdot & \cdot & \cdot \\
0 & \alpha_{n,1} & \cdot & \cdot & \alpha_{n,n-1}
\end{array}\right)\ .
\eeq{rescoup2}
It is easy to see that the group that preserves the form of these
matrices
is $\hat{G}=U(n)_Q\otimes U(n)_u\otimes U'(n-1)_{d}\otimes U(1)_B$
with the $U'(n-1)$ now acting on the $2,3,\cdots,n$ generations of
$d_R$ quarks, while the only symmetries of the full Lagrangian are
$U(1)_{d_1}\otimes U(1)_B$. Again we can do the counting
\beq
\begin{array}{||c|c|c||} \hline
\mrm{couplings and symmetries} & \mrm{moduli} & \mrm{phases} \\ \hline
(Y_u , Y_d)  &  n^2+n(n-1) & n^2+n(n-1)\\ \hline
U(n)_Q \otimes U(n)_u\otimes U'(n-1)_d\otimes U(1)_B
 & -(n-1)(3n-2)/2 & -n(3n+1)/2-1 \\ \hline
U(1)_{d_1}\otimes U(1)_B & 0 & 2 \\ \hline
 Y_{phys} &n+(n-1)+n(n-1)/2 & (n-1)(n-2)/2 \\ \hline
\end{array}
\eeq{compte3}
This model just gives a massless $d$-quark. Phases and mixings are
exactly the same as in the Standard Model for any number of generations.
This result is interesting because the so-called $m_u=0$ solution of
the Strong CP Problem (see for example \cite{Kim87,Che88,Pec89} and
references therein)  is based in a symmetry like ours.
In our
discussion we have obviated the fact that chiral symmetries have in
general anomalies. In the Standard Model, because of
QCD instanton effects
and because the chiral symmetries are anomalous, there is an additional
phase that cannot be removed. Chiral transformations move it from the
Yukawa sector to the QCD sector. This represents a problem because it
gives a too large contribution to the electric dipole moment of the
neutron  unless it is made unnaturally small.
If for instance $m_u=0$, there is an additional chiral symmetry
(similar to the one we have just studied) and the extra phase can be
rotated away.  However,  imposing an additional chiral symmetry is a very
strong constraint on the mass matrices [see \eq{rescoup2}]. One has to
redo all parameter counting and be sure that, there still is in the model
an observable CP-violating phase that could account for all observed
CP-violating effects. This is just what we did in our example.

The examples  discussed above show that the use of symmetry is
a natural and basis-independent way of choosing models for mass
matrices. Of course, most of the symmetries lead to some massless
quarks and/or some vanishing mixings, but the symmetries could
be softly broken in order to generate a realistic pattern of
masses and mixings. We find it interesting to explore and classify
all possible symmetries that can be imposed on the Yukawa sector
of the Standard Model according to the spectrum of masses and mixings
they generate. This is nothing but the classification of all
subgroups of $U(n)_Q\otimes U(n)_u\otimes U(n)_d$ according to the
spectrum they produce. We think that this classification is essential
for any mass-matrix modelling.

Let us apply our method of parameter counting to  a more
complicated example. Let us take an extension of the
Standard Model including right-handed neutrinos with a Majorana mass
term.
The leptonic part of the Lagrangian is
\beq
{\cal L } = i \overline{\ell_L} \db \ell_L
            +i \overline{\nu_R} \db \nu_R
            +i \overline{e_R} \db e_R
            + \left(\overline{\ell_L} Y_\nu \nu_R \varphi
            + \overline{\ell_L} Y_e e_R \tilde{\varphi}
            + \half \overline{\nu_R^c} M \nu_R\ +\ h.c. \right)
\eeq{lagrlept}
where  $\ell_L$ is the
standard leptonic doublet, $e_R$ is the right-handed
charged lepton singlet,  $\nu_R$ is the right-handed neutrino, and
$\nu_R^c = {\cal C} \overline{\nu_R}^T$.
The Yukawa couplings are the usual $Y_e$ that
give masses to the charged leptons,  $Y_\nu$ that give a
Dirac mass
term to neutrinos and  $M$ is a complex symmetric
Majorana mass matrix for right-handed
neutrinos. It can be put in
 by hand, as we did, because right-handed neutrinos
are singlet, however, it could be generated through the vacuum expectation
value of
some extra  scalar singlet. The model just described is nothing else
than the
{\it see-saw} model\cite{GR79,Yan79,MS80,MS81} of neutrino masses.
If $Y_\nu=Y_e=M=0$ the Lagrangian is invariant under the
following  symmetries
\beq
\ell_L \rightarrow V_\ell \ell_L\bla
\nu_R \rightarrow V_\nu \nu_R\bla
e_R \rightarrow V_e e_R\ ,
\eeq{trcampslept}
$V_\ell$, $V_\nu$, and $V_e$ are unitary matrices of dimension $n$ for
$n$ generations of leptons. As in the Standard Model,
the Yukawa couplings break these symmetries explicitly. However, they
are broken in such a way that if we let the Yukawa couplings and the
Majorana mass transform in the following way
\beq
Y_e \rightarrow V_\ell Y_e V_e^\dagger\bla
Y_\nu \rightarrow V_\ell Y_\nu V_\nu^\dagger\bla
M \rightarrow V_\nu^* M V_\nu^\dagger\  ,
\eeq{trmatriuslept}
the Lagrangian remains invariant.
At difference with the Standard Model, we see that the Yukawa
couplings and the Majorana mass term completely break all the invariances
of \eq{trcampslept}, including lepton-number conservation.

Equation (\ref{trmatriuslept})
 defines the following  equivalence relation
\beq
(Y_e,Y_\nu,M) \Leftrightarrow (Y'_e,Y'_\nu,M')=
(V_\ell Y_e V_\nu^\dagger,\ V_\ell Y_\nu V_e^\dagger,
V_\nu^* M V_\nu^\dagger )\  .
\eeq{releqlept}
Again the number of physical parameters needed to describe the model
is given by the difference between the number of parameters contained in
the Yukawa coupling and the number of generators broken by the
couplings.

Now the group of symmetries of the  kinetic and gauge part is
$G=U(n)_\ell\otimes U(n)_e \otimes U (n)_\nu$
and there is no unbroken subgroup.
Counting of parameters is easy. Taking into account
that an arbitrary complex
matrix of dimension $n$ contains $n^2$ moduli and $n^2$ phases,
that a symmetric complex matrix contains $n(n+1)/2$ moduli and
$n(n+1)/2$ phases, and that a $U(n)$
matrix
contains $n(n-1)/2$ moduli and $n(n+1)/2$ phases, we find that $Y_{phys}$
can be written in terms of
\beq
\begin{array}{||c|c|c||}\hline
\mrm{couplings and symmetries} & \mrm{moduli} & \mrm{phases} \\ \hline
(Y_e, Y_\nu, M)  &  2n^2+n(n+1)/2 & n^2+n(n+1)/2\\ \hline
U(n)_\ell \otimes U(n)_e\otimes U(n)_\nu
 & -3n(n-1)/2& -3n(n+1)/2\\ \hline
 & 0 & 0 \\ \hline
 Y_{phys} &3n+n(n-1) &  n(n-1) \\ \hline
\end{array}\ .
\eeq{compte4}

We obtain that even in the two-generation case there are
two observable phases that could violate CP. As we did in the Standard
Model, to know how many of the physical parameters are related to masses
and how many  to mixings,
we have to decouple all kinds of massive
bosons. Then, all the mass matrices can be diagonalized, leading
in general
to $n$ massive charged leptons and $2n$ Majorana neutrinos. Thus, $3n$ of
the moduli are masses, and the rest correspond to mixings.

To find a parametrization of the $Y_{phys}$,  we can
use the equivalence relation in \eq{releqlept} to reduce the number
of parameters in the Yukawa couplings. Using the fact
that an arbitrary complex matrix
can be written in a unique way as a unitary matrix times a
positive-definite Hermitian matrix\footnote{We will assume  that
the matrices have
non-zero determinant and there are no degeneracies. If there is some
degeneracy or zero eigenvalues, this must be imposed with some
symmetry and
would change the counting of parameters.}
and that a complex symmetric matrix
can be written in a unique way in terms of a
positive-definite diagonal matrix and a unitary matrix $S=U^T D U$ we can
write the Yukawa couplings in the following form
\beq
(Y_e,Y_\nu,M) \rightarrow Y_{phys} =(H_e , H_\nu , D_m)\  .
\eeq{paramlept}
Here $H_e$ and $H_\nu$ are  positive-definite Hermitian
matrices and $D_m$ is a positive-definite diagonal matrix.
As we have exhausted all the symmetries in the group $G$ to
reduce the Yukawa couplings to this form and there
is no subgroup of $G$ that leaves $Y_{phys}$ invariant,  the set of
parameters in the right-hand side of \eq{paramlept} is the
physical set of parameters needed to describe the theory.
This means
that we could start from the beginning without loss of generality by
choosing $Y_\nu$ and $Y_e$ as
positive-definite Hermitian matrices and $M$ as a positive
definite diagonal matrix and {\it all} the parameters in
those matrices would be
observable as  can be easily checked by counting the number of parameters
contained in those couplings.

Finally to give a flavour of how these techniques can
be used to study more exotic
models, we chose the model of refs.
\cite{BL77,CL77,WW83,BS87} with the following
Lagrangian,
\[
{\cal L } = i \overline{\ell_L} \db \ell_L
            +i \overline{\nu_R} \db \nu_R
            +i \overline{e_R} \db e_R
            +i \overline{s_L} \db s_L
\]
\beq
            + \left(\overline{\ell_L} Y_\nu \nu_R \varphi
            + \overline{\ell_L} Y_e e_R \tilde{\varphi}
            + \half \overline{\nu_R} M_s s_L\ +\ h.c. \right) \   ,
\eeq{lagrlept2}
where $s_L$ is a singlet that carries the same lepton-number as
the leptons
 and $M_s$ is an arbitrary complex matrix. The rest of the notation is the
 same as  in the previous example.
In the Lagrangian of \eq{lagrlept2},
total lepton-number has been imposed as
a global symmetry. There are thus no Majorana mass terms for any of
the singlet fermions. The group
of invariances of the non-Yukawa part of the Lagrangian is
$U(n)_\ell \otimes U(n)_e \otimes U(n)_\nu \otimes U(n)_s$,
where $U(n)_s$ is the new invariance related to the field $s_L$.
 The only symmetry of the Yukawa couplings is just $U(1)_{lep}$
 of lepton-number conservation.
 Doing now the usual counting we get
\beq
\begin{array}{||c|c|c||}\hline
\mrm{couplings and symmetries} & \mrm{moduli} & \mrm{phases} \\ \hline
(Y_e , Y_\nu  , M_s ) & 3n^2 & 3n^2 \\ \hline
 U (n)_\ell \otimes U(n)_e \otimes U(n)_\nu\otimes U(n)_s &
 -4n(n-1)/2 & -4 n(n+1)/2 \\ \hline
 U(1)_{lep}  & 0 & 1 \\ \hline
  Y_{phys} & 2n+ n^2  & (n-1)^2  \\ \hline
\end{array} \ .
\eeq{brancovalle}
Once the gauge symmetry is broken,
the Yukawa couplings generate mass terms
for the leptons.
It is easy to see that, as a consequence of lepton-number
conservation, all neutrinos must be  Dirac neutrinos or just massless.
Then, from the full mass matrix of neutrinos, $n$ massless neutrinos
and $n$ massive Dirac neutrinos arise, hence only $n$
masses come from the
neutrino mass matrix.  The other $n$ masses come from the charged lepton
sector.  Thus, from all the moduli, $2n$ of them correspond to masses
and the rest $n^2$  correspond to mixings.  In addition  there are
$(n-1)^2$  CP-violating phases  that cannot be removed.
 This result is in complete
agreement with the results obtained in \cite{BR89}
after full diagonalization
of the mass matrices.

We have shown that a study of the chiral symmetries of the kinetic and
gauge parts
of the Lagrangian and the Yukawa part of the
Lagrangian before spontaneous
symmetry breaking allows us
to compute,
in a straightforward way, how many observable parameters come from
the Yukawa couplings in a general gauge theory.
Basically, we have to know which subgroup of chiral symmetries of the
kinetic and gauge parts of the Lagrangian preserves
the form of the Yukawa
couplings and which group leaves them invariant. The number of
parameters that can be absorbed from  the Yukawa sector is just
the difference
between the parameters needed to describe those groups. Then,
the number of observable parameters
comes from the balance between the Yukawa couplings and
the symmetries broken by them. The method is useful when the
diagonalization of the full
set of mass matrices becomes complicated. On the
other hand it clarifies the analysis of observable parameters.
We explain the technique with several examples based on the Standard
Model, with Yukawa couplings obeying various chiral symmetries, and also
with some extensions of the Standard Model with enlarged fermion sector.

The study of the mass matrices and the gauge couplings
is not general enough,
because some parameters such as phases,
mixings, etc, can be moved from charged
currents to Yukawa couplings or even be included in a field definition
\cite{BP83,BP86}. Only the analysis of the full Lagrangian, and this is
better done before symmetry breaking, can report the number of observable
parameters in a non-ambiguous way.
This point of view suggests a parametrization
of Yukawa couplings in terms of invariants with
respect to the chiral symmetries
of the kinetic and gauge parts of the Lagrangian
that can be useful in the study of the
behaviour of Yukawa couplings under the renormalization group. We comment
on this parametrization in the case of the Standard Model and the see-saw
model for neutrino masses.

We find that a complete analysis
and classification of all the
chiral symmetries that can be imposed on the
Yukawa sector, according to which spectrum  of masses and mixings
they originate is still missing. We think this analysis would be a good
starting point for mass matrix modeling.

Finally we want to remark that we only considered symmetries acting
 on fermions. The analysis could
 be extended to symmetries involving fermions and
scalars as well, like
the Peccei-Quinn \cite{PQ77,PQ77a}  symmetry or horizontal symmetries
\cite{GM79,SW79,WZ79,GM80}.
We also think that it can be extended to the spontaneous
CP-violation case.

\section*{Acknowledgments}
We thank F. Botella for helpful discussions
on the subject of this paper and A. Pich for a critical reading
of the manuscript. This work has been supported in part by
CICYT, Spain, under grant AEN90-0040.

\end{document}